\documentclass[%
 reprint,
superscriptaddress,
 amsmath,amssymb,
prb,
]{revtex4-2}

\usepackage{graphicx}
\usepackage{dcolumn}
\usepackage{bm}
\usepackage{comment}
\usepackage{xcolor}

\begin{document}

\preprint{APS/123-QED}

\title{Axion-Field-Enabled Nonreciprocal Thermal Radiation in Weyl Semimetals}

\author{Bo Zhao}\thanks{These authors contributed equally.}
\affiliation{%
 Ginzton Laboratory and Department of Electrical Engineering, Stanford University, Stanford, California 94305, USA
}

\author{Cheng Guo} \thanks{These authors contributed equally.}
 \affiliation{Department of Applied Physics, Stanford University, Stanford, California 94305, USA}

\author{Christina A.~C.~ Garcia}
\affiliation{John A.~Paulson School of Engineering and Applied Sciences, Harvard University, Cambridge,Massachusetts 02138, USA}

\author{Prineha Narang}
\affiliation{John A.~Paulson School of Engineering and Applied Sciences, Harvard University, Cambridge,Massachusetts 02138, USA}

\author{Shanhui Fan}
\email{shanhui@stanford.edu}
\affiliation{%
 Ginzton Laboratory and Department of Electrical Engineering, Stanford University, Stanford, California 94305, USA
}

\date{\today}

\begin{abstract}

Objects around us constantly emit and absorb thermal radiation \cite{kirchhoff1860,planck1989,howell2016,bergman2011,chen2005,zhang2007}. The emission and absorption processes are governed by two fundamental radiative properties: emissivity and absorptivity. For reciprocal systems, the emissivity and absorptivity are restricted to be equal by Kirchhoff's law of thermal radiation  \cite{kirchhoff1860,planck1989,landau2008}. This restriction limits the degree of freedom to control thermal radiation and contributes to an intrinsic loss mechanism in photonic energy harvesting systems such as solar cells \cite{green2012}. Existing approaches to violate Kirchhoff's law typically utilize conventional magneto-optical effects in the presence of an external magnetic field \cite{Zhu2014a,Zhao2019d}. However, these approaches require either a strong magnetic field (${\sim}3$T) \cite{Zhu2014a}, or narrow-band resonances under a moderate magnetic field (${\sim}0.3$T) \cite{Zhao2019d}, because the non-reciprocity in conventional magneto-optical effects  is usually weak in the thermal wavelength range. Here, we show that the axion electrodynamics in magnetic Weyl semimetals can be used to construct strongly nonreciprocal thermal emitters that near completely violate Kirchhoff’s law over broad angular and frequency ranges, without requiring any external magnetic field. The non-reciprocity moreover is strongly temperature tunable, open new possibilities for active non-reciprocal devices in controlling thermal radiation.   
\end{abstract}

\maketitle



Most thermal emitters obey Kirchhoff’s law, which states that the emissivity and the absorptivity are equal for a given direction, frequency, and polarization \cite{kirchhoff1860}. However, Kirchhoff's law is not the requirement of  thermodynamic laws but rather results from Lorentz reciprocity \cite{landau2008,fan2017}. Therefore, Kirchhoff's law can be broken by nonreciprocal emitters made of e.g.~gyrotropic materials with asymmetric permittivity tensors  \cite{ries1983, Miller2017}. Such nonreciprocal emitters open up the significant possibility to control  emission and absorption separately, which can  enable photonic heat engines to reach their ultimate thermodynamic limit \cite{landsberg1980,green2012}. 

Despite the profound implications of nonreciprocity in thermal radiation, there are only a limited number of proposed nonreciprocal thermal emitters that can achieve significant violation of Kirchhoff's law, primarily due to the lack of materials that exhibit strong nonreciprocal response in mid-infrared. The difference in the  emissivity and absorptivity depends on the the degree of asymmetry of the permittivity tensor $\overline{\overline{\varepsilon}}$, which can be characterized by $\gamma = \|\overline{\overline{\varepsilon^A}}\|/\|\overline{\overline{\varepsilon^S}}\|$, where $\|\cdot\|$ is the matrix norm, and $\overline{\overline{\varepsilon^S}}$ and $\overline{\overline{\varepsilon^A}}$ are the symmetric and antisymmetric part of $\overline{\overline{\varepsilon}}$, respectively. For existing nonreciprocal emitters made of magneto-optical materials, $\gamma\sim \omega_c /\omega$, where $\omega_c = eB/m^*$  is the cyclotron frequency, $\omega$ is the radiation frequency \cite{Zhu2014a},  $m^*$ is the effective electron mass, and $B$ is the external magnetic field. For a typical magnetic field $B\sim1$T, $\omega_c\sim1$THz, thus $\gamma\sim 0.01$ in mid-infrared. Such a weak nonreciprocity explains why in previous works either a large magnetic field or a high quality-factor  (Q-factor) resonance is required to achieve significant violation of Kirchhoff's law \cite{Zhu2014a, Zhao2019d}.

\begin{figure}[htbp]
\includegraphics[width=1.0\columnwidth]{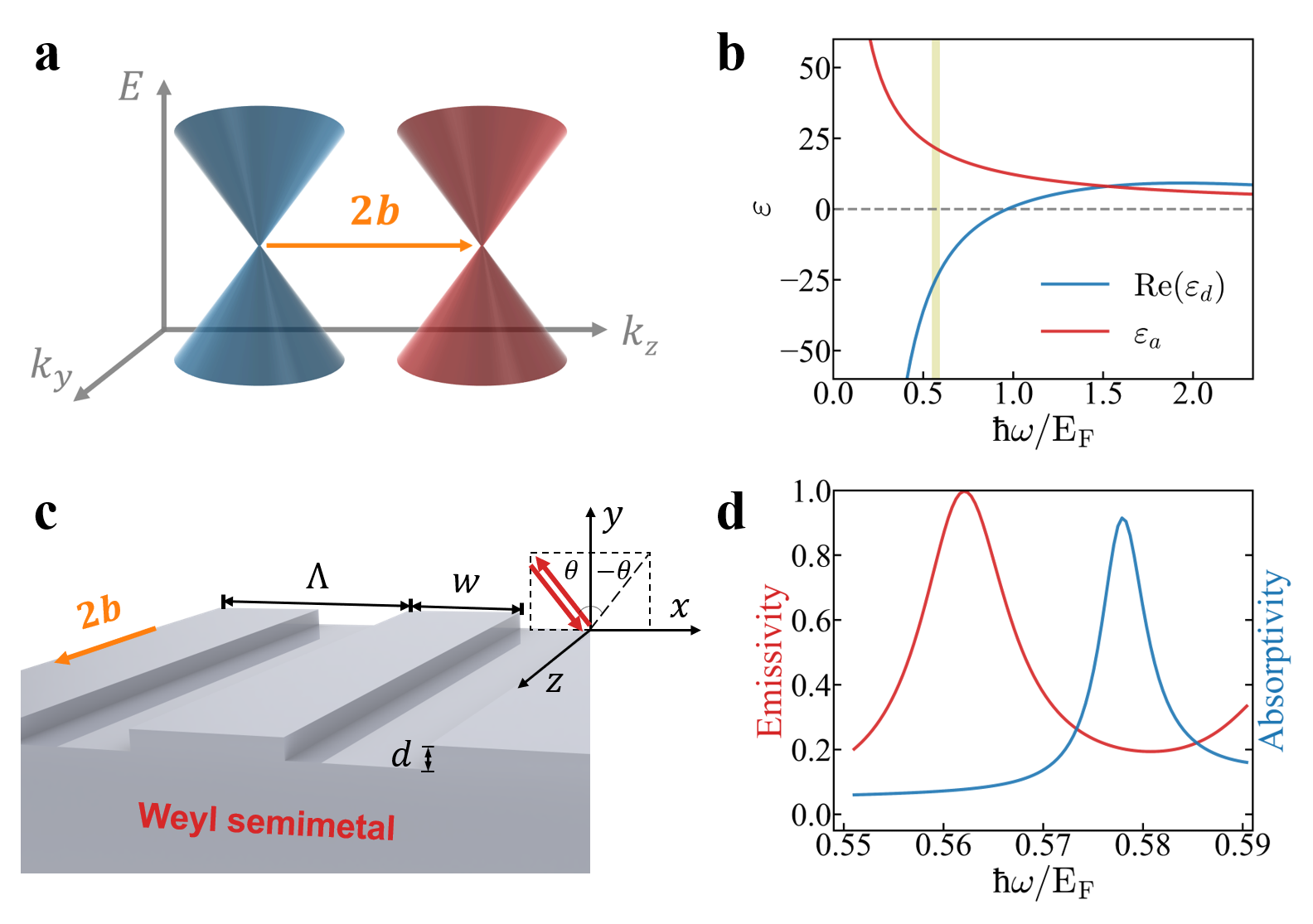}
\centering
\caption{(a) Electronic band structure of a magnetic Weyl semimetal with two Weyl nodes of opposite chirality separated by $2\mathbf{b}$ in momentum space. (b) The permittivity tensor components of the Weyl semimetal. The frequency $\omega$ is normalized by the  chemical potential $E_F$ at $T=300$~K. $\varepsilon_d$ and $\varepsilon_a$ are the diagonal and off-diagonal component, respectively. The beige region highlights the frequency range used for thermal emitter design. (c) Schematic of a Weyl semimetal photonic crystal for  nonreciprocal thermal emission. The structure is purely made of the Weyl semimetal with $\mathbf{b}$ along the $z$ direction. It consists of a grating structure atop an optically-thick substrate. The structure is periodic in the $x$ direction with the following geometric parameters: $\Lambda = 7\,\mu$m, $w = 3.5\,\mu$m, $d = 0.5\,\mu$m. TM polarization with an electric field in the $x$-$y$ plane is considered. (d) Emissivity and absorptivity spectrum in $\theta = 80^\circ$ direction. 
} 
\label{fig:fig1}
\end{figure}

In this paper, we propose to use topological magnetic Weyl semimtals to construct non-reciprocal thermal emitters.  (Fig.~\ref{fig:fig1}(a)). Magnetic Weyl semimetals can exhibit highly unusual and extremely-large gyrotropic optical responses with $\gamma\sim1$ in the mid-infrared, which originates from their unique topologically nontrivial electronic states and inherent time-reversal symmetry breaking. (Fig.~\ref{fig:fig1}(b)). Based on this strong nonreciprocity, we design a nonreciprocal thermal emitter (Fig.~\ref{fig:fig1}(c)) that can provide near-complete violation of the Kirchhoff's law at the wavelength of  around 15~$\mu$m (Fig.~\ref{fig:fig1}(d)), without the need for an external magnetic field. We show that the nonreciprocal thermal properties are highly temperature dependent because of the strong temperature dependence of the chemical potential. Although the nonreciprocal optical response of a magnetic Weyl semimetal is already predicted \cite{Kotov2018}, its implication in thermal radiation has not been explored before.

Weyl semimetals are a newly discovered class of three-dimensional gapless topological matter \cite{armitage2018a,yan2017,Hosur2013}. They feature accidental degenerate points in their band structure — called the Weyl nodes — that host chiral fermions and appear in pairs of opposite chirality. Each Weyl node acts as a source/drain of Berry curvature in momentum space. Realization of a Weyl semimetal requires breaking of either inversion ($\mathcal{P}$) or time-reversal ($\mathcal{T}$) symmetry. Noncentrosymmetric Weyl semimetals that break $\mathcal{P}$ but preserve $\mathcal{T}$ symmetry have been discovered since 2015  \cite{Xu2015,Lv2015}, but only until more recently have magnetic Weyl semimetals that break $\mathcal{T}$ been experimentally discovered   \cite{belopolski2019,morali2019,liu2019,kuroda2017,hirschberger2016}. These magnetic Weyl semimetals have generated significant interest both for their fundamental physics properties, and for their potential applications such as in spintronics due to their intrinsic magnetism and large observed anomalous Hall conductivity  \cite{Liu2018,Wang2018} as well as in spin caloritronics  because of their giant anomalous Nernst coefficient  \cite{Sakai2018}.

The nontrivial topology of the Weyl nodes gives rise to novel electromagnetic responses that are drastically different from conventional materials. These include the chiral magnetic effect and the anomalous Hall effect \cite{armitage2018a}. Both effects can be represented compactly by the formalism of axion electrodynamics \cite{wilczek1987}, which adds the axion term to the electromagnetic Lagrangian density \cite{zyuzin2012}:
\begin{align}
\label{eq:lagrangian}
    \mathcal{L}_\theta &= 2\alpha \sqrt{\frac{\epsilon_0}{\mu_0}}\frac{\theta(\mathbf{r},t)}{2\pi}\mathbf{E}\cdot \mathbf{B},
\end{align}
where  
$\alpha = \tfrac{e^2}{4\pi\epsilon_0\hbar c}$ is the fine structure constant, $e$ is the elementary charge, $\hbar$ is the reduced Planck constant, $\epsilon_0$ is the permittivity of vacuum, $\mu_0$ is the permeability of vacuum, $\mathbf{E}$ is the electric field, $\mathbf{B}$ is the magnetic flux density, and $\theta(\mathbf{r},t)$ is the axion angle that has space and time dependence. 

For concreteness, we consider the simplest case of a magnetic Weyl semimetal with two Weyl nodes of opposite chirality. Such an ideal Weyl semimetal has been experimentally realized very recently \cite{soh2019}. In this case,
\begin{equation}
\label{eq:theta}
    \theta(\mathbf{r},t) = 2(\mathbf{b}\cdot\mathbf{r}-b_0t),
\end{equation}
where $2\hbar\mathbf{b}$ and $2\hbar b_0$ are the separation of the two Weyl nodes in momentum and energy, respectively. Such an axion term results in the modified constitutive relations in the frequency domain \cite{Hofmann2016}:
\begin{equation} \label{eq:constitutive}
\mathbf{D} = \varepsilon_d \mathbf{E} + \frac { i e ^ { 2 } } {  4 \pi^2  \hbar \omega } ( -2b_0 \mathbf{B} + 2\mathbf{b} \times \mathbf{E} ) 
\end{equation}
Here $\varepsilon_d$ is the permittivity of the corresponding Dirac semimetal which has doubly degenerate bands with $b_0 = \mathbf{b} = 0$. Below    we assume $\epsilon_d$ is isotropic. The first and the second terms in the parentheses describe the chiral magnetic effect and the anomalous Hall effect, respectively. In this paper we consider only materials where the  Weyl nodes have the same energy (i.e., $b_0 = 0$). The momentum-separation $2\mathbf{b}$ of the Weyl nodes is an axial vector that acts similar to an internal magnetic field, and we choose the coordinates such that $\mathbf{b}$ is along the $k_z$ direction: $\mathbf{b} = b \mathbf{\hat{k}_z}$.  (Fig.~\ref{fig:fig1}(a)). With the above considerations, the permittivity tensor of the Weyl semimetal becomes:

\begin{equation}\label{eq:epsilon_tensor}
\overline{\overline{\varepsilon}} =\begin{bmatrix}
\varepsilon_d&i\varepsilon_a&0\\
-i\varepsilon_a&\varepsilon_d&0\\
0&0&\varepsilon_d
\end{bmatrix},
\end{equation}
where 
\begin{equation}\label{eq:epsilon_a}
\varepsilon _ { a } = \frac { b e ^ { 2 } } { 2 \pi^2 \hbar \omega }.
\end{equation}
Since $b\neq 0$, $\epsilon_{a}$ becomes nonzero, therefore $\overline{\overline{\varepsilon}}$ is asymmetric and breaks Lorentz reciprocity.  Such nonreciprocity is due to the intrinsic anomalous Hall effect induced by the Berry curvature associated with the Weyl nodes, which  fundamentally differ from the cyclotron mechanism in magneto-optical materials in that no external magnetic field is required.

To calculate the diagonal term $\varepsilon_d$, we apply the Kubo-Greenwood formalism within the random phase approximation to a two-band model with spin degeneracy \cite{Hofmann2016,Kotov2016,Kotov2018}. This formalism takes into account both interband and intraband transitions:
\begin{equation}\label{eq:ed}
    \varepsilon_d = \varepsilon _ { b } +  i \frac { \sigma  } { \omega }
\end{equation}
Here $\sigma$ is the bulk conductivity and given by
\begin{align}
    \label{eq:sigma}
    \sigma = \frac { r _ { s } g } { 6 } \Omega \, G ( \Omega / 2 ) + 
    i \frac { r _ { s } g } { 6 \pi  } \left\{ \frac { 4 } { \Omega } \left[ 1 + \frac { \pi ^ { 2 } } { 3 } \left( \frac { k _ { B } T } { E _ { F } (T)} \right) ^ { 2 } \right] \right.\notag\\
    + \left. 8 \Omega \int _ { 0 } ^ { \xi _ { c } } \frac { G ( \xi ) - G ( \Omega / 2 ) } { \Omega ^ { 2 } - 4 \xi ^ { 2 } } \xi d \xi \right\}\,,
\end{align} 
$\varepsilon_b$ is the background permittivity, $\Omega = \hbar(\omega+i\tau^{-1})/E_F$  is the complex frequency normalized by the chemical potential, $\tau^{-1}$ is the scattering rate corresponding to Drude damping, $G(E) = n(-E)-n(E)$,  where $n(E)$ is the Fermi distribution function, $E_F(T)$ is the chemical potential, $r_s = e^2/4\pi\epsilon_0\hbar v_F$ is the effective fine structure constant, $v_F$ is the Fermi velocity, $g$ is the number of Weyl points, and $\xi_c =E_c/E_F$ where $E_c$ is the cutoff energy beyond which the band dispersion is no longer linear \cite{Kotov2016}. Following Ref.~[ \citenum{Kotov2018}], in this work, we use the parameters $\varepsilon_b = 6.2$, $\xi_c = 3$, $\tau= 1000$~fs, $g=2$, and $v_F = 0.83 \times 10^5$~m/s. We choose the chemical potential to be $E_F = 0.15$~eV at $T=300$~K, which is typical for doped Weyl semimetals. We have verified that the permittivity tensor calculated from this model is consistent with that calculated from another effective Hamiltonian approach \cite{Chen2013}. In Fig.\ref{fig:fig1}(b), we plot $\varepsilon_d$ and $\varepsilon_a$ at $T=300$~K. We see that the off-diagonal matrix element $\varepsilon_a$ is comparable to the $\epsilon_d$ in the whole displayed wavelength range, indicating a strength of non-reciprocity significantly exceeding that of conventional  magneto-optical materials.

In Fig.~\ref{fig:fig1}(c), we show the schematic of our proposed nonreciprocal emitter, consisting of  a photonic crystal  made of Weyl semimetal. The photonic crystal possesses a periodic array of strips along the $x$ direction. The period is $\Lambda = 7 \mu$m and the strip width is $w = 3.5 \mu$m. The thickness of the grating is $d = 0.5 \mu$m, and the grating is atop an optically-thick substrate. Therefore the structure is opaque and the transmission is zero. The Weyl semimetal is arranged in the Voigt configuration, i.e. $\mathbf{b}$ is along the $z$ direction and the plane-of-incidence is the $xy$-plane. The angle of incidence (or emission) is denoted as $\theta$. We consider the transverse magnetic (TM) waves with the electric field in the $xy$-plane, as this is the polarization that experiences the nonreciprocal effect. 

With this configuration, the emitter exhibits only  specular reflection in the wavelength range of interest, i.e.~a plane  wave with an angle of incidence  $\theta$ is either absorbed by the structure or reflected to the complementary channel in the $-\theta$ direction. The former process yields an absorptivity of $\alpha_\theta$ whereas the latter yields a reflectivity $R_\theta$. From energy balance, $\alpha_\theta = 1- R_\theta$. On the other hand, the emissivity of channel $\theta$ can be related to the reflectivity of the complementary channel by the thermodynamic argument presented in Refs.~[ \citenum{Zhu2014a,Zhao2019d}], which states that $\epsilon_\theta = 1- R_{-\theta} = \alpha_{-\theta}$. For nonreciprocal emitters, $R_{\theta} \neq R_{-\theta}$ and therefore $\alpha_{\theta}$ $\neq$ $\epsilon_{\theta}$. In Fig.~\ref{fig:fig1}(d) we show $\epsilon_{80^\circ}$ and $\alpha_{80^\circ}$ as functions of $\omega$ for TM polarization. The large contrast between the emissivity and absorptivity, especially near the peak frequency of the emissivity spectrum ($\omega\approx0.56 E_F/\hbar$), indicates a near complete violation of Kirchhoff's law.

\begin{figure}[htbp]
\includegraphics[width=1.0\columnwidth]{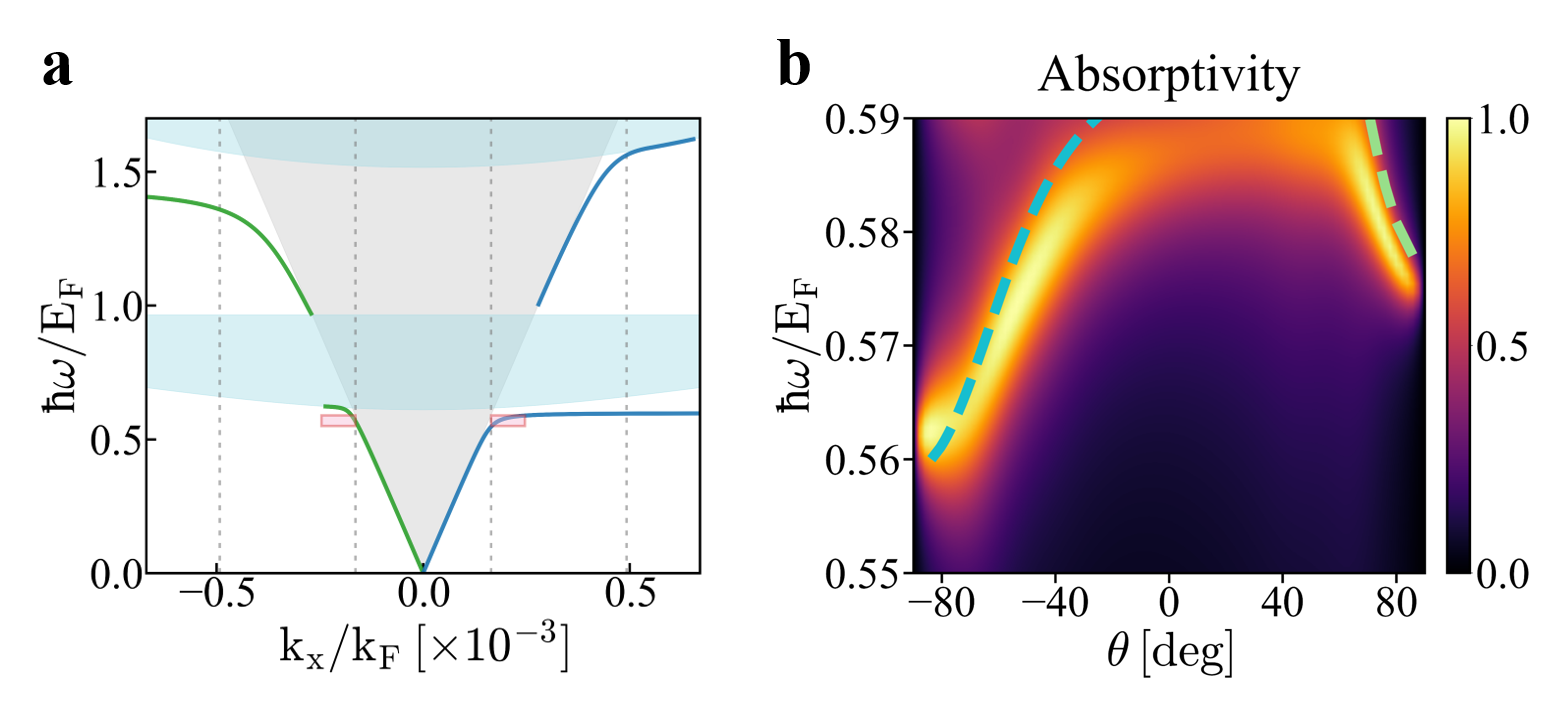}
\centering
\caption{(a) Nonreciprocal surface plasmon dispersion of the Weyl semimetal. The gray region shows the light cone of vacuum. The blue region shows  the continuum of bulk modes. The red region highlights the asymmetric surface plasmon dispersion used for thermal emitter design. The gray dashed lines show the boundaries of the Brillouin zones. $k_F = E_F/\hbar v_F$ is the Fermi wavevector. (b) The angular and frequency dependence of  absorptivity. The dashed lines are predictions from the folded  dispersion of the surface plasmon polariton. 
}
\label{fig:fig2}
\end{figure}

Since $\epsilon_\theta - \alpha_\theta  = R_\theta - R_{-\theta}$, one needs to maximize $|R_{\theta} - R_{-\theta}|$ to achieve a strong violation of Kirchhoff's law. In the design above, we  achieve this by critically coupling the incident wave to one branch of the asymmetric surface plasmon polariton modes in the Weyl semimetal photonic crystal. The band dispersion of the photonic crystal can be understood by first considering a semi-infinite Weyl semimetal.  Its dispersion relation is calculated from
\begin{equation}\label{eq:dispersion}
k _ { z } +\varepsilon _ { v}  k _ {z0 }  - k _ { x }\dfrac{\varepsilon _ { a }} {\varepsilon _ { d }}= 0\,.
\end{equation}
where $k_0 = \omega/c$, $k_x$ is the parallel wave vector of the surface plasmon polaritons, and $k_{z0} = \sqrt{k_x^2- k_0^2}$ and $k_z= \sqrt{k_x^2-\epsilon_v k_0^2}$ are the $z$-component of the wave vector in air and in the Weyl semimetal, respectively, and  $\varepsilon_v=\varepsilon_d-\varepsilon_a^2/\varepsilon_d$ . The last term in the above equation shows that when the sign of $k_x$ changes, the frequency of the surface plasmon polaritons will be different. In Fig.~\ref{fig:fig2}(a), we show the calculated band dispersion, together with the continuum region of bulk modes (in light blue) and the light cone of vacuum (in light gray). We can clearly observe the asymmetric band structures of the surface plasmon polaritons outside the light cone and the continuum region of bulk modes. In Fig.~2(a) we also highlight in red the region of the asymmetric surface plasmon dispersion that is used for thermal emitter design.

To excite the surface plasmon polariton wave  with a free-space optical wave, in the design shown in Fig.~\ref{fig:fig1}(c)  we use a grating with  a periodicity $\Lambda = 7$ $\mu $m to compensate the wavevector mismatches between the two waves. In Fig.~\ref{fig:fig2}(a), we draw the boundaries of the resultant Brillouin zones in vertical dashed lines.
In our design, the incident wave at $\omega = 0.56 E_F/\hbar$ ($\lambda = 14.7$~$\mu$m) from $\theta = - 80^\circ$ can critically couple to the surface plasmon polariton, which yields $R_{-80^\circ}=0$ and thus $\epsilon_{80^\circ}=1$, while the incident light at the same frequency from $\theta = 80^\circ$ has no surface plasmon mode to couple to, which yields $R_{80^\circ}\approx1$ and thus $\alpha_{80^\circ}\approx0$. Therefore, the asymmetric band structure of the surface plasmon polaritons leads to the large difference in the emissivity and absorptivity.

Our design supports the violation of Kirchhoff's law in a wide angular and frequency range. In Fig.~\ref{fig:fig2}(b) we plot the absorptivity $\alpha(\omega, \theta)$. The emissivity can be inferred as $\epsilon(\omega, \theta) = \alpha(\omega, -\theta)$. In the frequency range $0.56<\hbar\omega/E_F<0.575$, the absorptivity is particularly strong for $\theta<0$, and thus emissivity is particularly strong for $\theta>0$. Therefore, the absorption and emission processes are effectively decoupled. Fig.~\ref{fig:fig2}(b) also plots the folded band dispersion of the surface plasmon polaritons in dashed lines, which agrees well with the absorption peaks. This confirms that the nonreciprocal thermal properties originates from the nonreciprocal surface plasmon polaritons in the Weyl semimetal.

\begin{figure}[htbp]
\includegraphics[width=1.0\columnwidth]{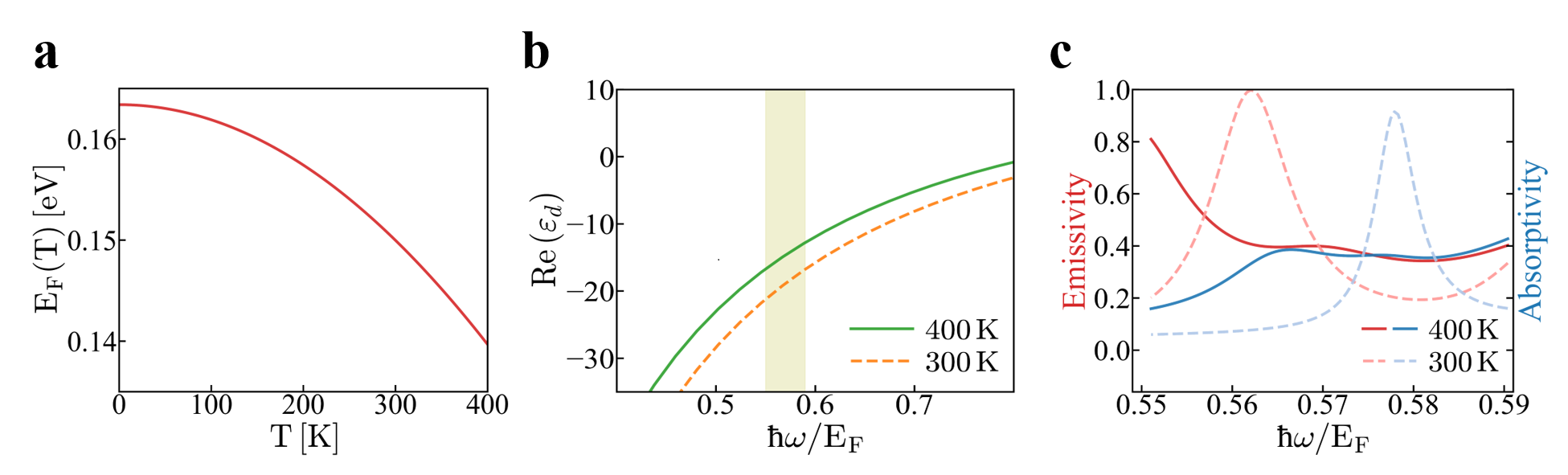}
\centering
\caption{(a) The temperature dependence of the chemical potential $E_F(T)$. (b) $\operatorname{Re}(\epsilon_d)$ at $T=300$~K and $T=400$~K. The beige region highlights the  frequency range used for  thermal emitter design. (c) Emissivity and absorptivity in $\theta = 80^\circ$ direction at $T=300$~K and $T=400$~K for the structure shown in Fig. 1(c). 
}
\label{fig:fig3}
\end{figure}

In the above discussions, we assume the emitter is at room temperature ($T=300$~K). For usual thermal emitters, it is typical to neglect the temperature dependence of the material properties. However, this does not apply to thermal emitters based on Weyl semimetals. A Weyl semimetal has a small and nonconstant density of states due to its linear dispersion, which causes its chemical potential to be strongly temperature dependent  \cite{Ashby2014}. Consequently, its  dielectric tensor and thermal properties are also temperature dependent. The chemical potential as a function of temperature can be calculated from charge conservation \cite{Ashby2014}:
\begin{equation}
\resizebox{1.0\columnwidth}{!}{$
E_F ( T ) = \frac { 2 ^ { 1 / 3 } [ 9 E_F (0) ^ { 3 } + \sqrt { 81 E_F(0) ^ { 6 } + 12 \pi ^ { 6 } k_B^6 T ^ { 6 } } ] ^ { 2 / 3 } - 2 \pi ^ { 2 } 3 ^ { 1 / 3 } k_B^2 T^2 } { 6 ^ { 2 / 3 } [ 9 E_F(0)^ { 3 } + \sqrt { 81 E_F(0)^ { 6 } + 12 \pi ^ { 6 } k_B^6 T ^ { 6 } } ] ^ { 1 / 3 } }$}
\end{equation}
where $E_F (0)= 0.163 $~eV such that $E_F (300$K$)= 0.150$~eV. In Fig.~\ref{fig:fig3}(a) we plot $E_F(T)$, which decreases as $T$ increases. Consequently $\varepsilon_d$ in Eq.~(\ref{eq:epsilon_tensor}) also exhibits strong temperature dependence. As an example, we observe in Fig.~\ref{fig:fig3}(b) a significant difference in the spectrum of $\operatorname{Re}(\varepsilon_d)$, between the cases with $T = 300$~K and $400$~K, as calculated using Eqs.~(\ref{eq:ed}) and (\ref{eq:sigma}).  
At these two temperatures the surface plasmon frequencies are also quite different, which manifests in the absorptivity and emissivity spectra.  
Fig.~\ref{fig:fig3}(c) shows $\alpha_{80^\circ}$ and $\epsilon_{80^\circ}$ as functions of $\omega$ at $T=300$~K (dashed lines, the same as Fig.\ref{fig:fig1}(d)) and $400$~K (solid lines). The contrast of absorptivity and emissivity is significantly reduced at $400$~K. Therefore, the nonreciprocal effect in the designed thermal emitter is very temperature sensitive. Such a strong temperature dependence of nonreciprocity may be used to switch the nonreciprocal thermal emission on and off for a certain frequency range, or construct active thermal devices  such as thermal rectifiers \cite{otey2010}.

In this work, we focus on the frequency and wavevector regime where the nonreciprocal surface plasmon modes are associated with the collective behavior of bulk carriers close to the surface. Weyl semimetals also feature topological Fermi arc surface states \cite{belopolski2016}, which can also support surface plasmon modes \cite{song2017}. The study of nonreciprocal thermal emission from such Fermi arc plasmons will be presented in our future work.

In conclusion, we demonstrate that the axion electrodynamics in topological magnetic Weyl semimetals provides a novel and effective mechanism to create nonreciprocal thermal emitters. We propose a Weyl-semimetal-based photonic crystal design that can achieve near-complete violation of Kirchhoff’s law without any external magnetic field. The violation of Kirchhoff’s law persists over a wide angular and frequency range, and the nonreciprocal radiative properties are highly  temperature sensitive. Our work puts forth topological quantum  materials as a solution to the long-standing materials challenge in constructing  nonreciprocal thermal emitters of practical interests. The designed Weyl-semimetal-based nonreciprocal thermal emitters could enable exciting opportunities in energy harvesting and heat transfer applications.

\section*{Acknowledgement}

This work was supported by the DOE Photonics at Thermodynamic Limits Energy Frontier Research Center under Grant No. DE-SC0019140.
C.A.C.G. is supported by the NSF Graduate Research Fellowship Program under Grant No. DGE-1745303.
P.N. is a Moore Inventor Fellow and a CIFAR Azrieli Global Scholar. The authors also acknowledge discussions with Yoichiro Tsurimaki, Dr.~Xin Qian, 
Simo Pajovic and Prof. Chen Gang, who pointed out their related unpublished works. 

%

\end{document}